\documentclass{article}
\usepackage{arxiv}
\usepackage{graphics}
\usepackage{graphicx}
\usepackage{pgfplots}
\usepackage{subcaption}

\pgfplotsset{compat=1.18}
\AtBeginDocument{%
  }

\usepackage{listings}
\begin{document}
\title{New simple and fast  quicksort algorithm for equal keys

 }
 \author { parviz afereidoon\\
\texttt{afereidoon.p@gmail.com}
}

\begin {abstract}
This paper introduces a  novel and efficient partitioning technique for quicksort, specifically designed for real-world data with duplicate elements (50-year-old problem). The method is referred to as "equal quicksort" or "eqsort". Based on the experimental findings, it has been determined that the newly developed algorithm, eqsort, is competitive with the best current implementations,such as fat partitioning algorithms and dual-pivot quicksort. This method offers several advantages over the commonly used dual-pivot method and pdqsort partitioning, making it a potential replacement.\\
\\
\\
\end {abstract}

\maketitle

\section {Introduction}
Data sorting is one of the fundamental issues in computer science. One of the most famous, most practical and fastest methods for sorting primitive data is Quicksort.   Quicksort is an efficient algorithm that was developed by tony hoare  in 1960 \cite{hoa},\cite{HO}. It is still a commonly used algorithm for sorting and  faster than other methods for  randomized data. Quicksort is a divide-and-conquer algorithm. It works by selecting a pivot element from the array and partitioning the other elements into two sub-arrays, according to wether they are less than or greater than the pivot, the sub-arrays are then sorted recursively.\\
Ian munro and philip spira for the first time in 1976 consider the problem of sorting a multiset (data with equal keys) \cite{MS}. One year later Robert Sedgewick  studied some algorithms  for handling  equal elements \cite{Rs} and he concluded that  ”the method of always stopping the scanning pointers on keys equal to the partitioning element performs best" . In this Sedgewick's partitioning, the elements larger than the partition key are placed on the right side,  the elements smaller than the  partition key are placed on the left side and equal elements are placed on both sides (equal elements of each side are transferred to the other side).
\\
In subsequent studies, alongside single-pivot 2-way partitioning techniques, researchers also identified and examined single-pivot 3-way partitioning methods, also known as fat partitioning and the multi-pivot partitioning methods as potential solutions for the issue of duplicated elements\cite{KM}. In the subsequent section, we outline several significant methodologies:
\\
\begin{itemize}
\item {\texttt{The Dijkstra partitioning method}}: This method consists of three parts. The left partition includes elements that are smaller than the partition key. The middle partition includes all elements that are equal to the partition key. Lastly, the right partition includes all elements that are greater than the partitioning key \cite{Dj}.\\
\item {\texttt{The Bentley-Mcllroy partitioning method}}: In 1993, John Bentley and Douglas Macllroy developed a quicker algorithm for handling equal elements compared to previous methods \cite{Bm}. This method bears resemblance to Dijkstra's partitioning, with the distinction that the elements at the extreme ends are equivalent to the partition key. Following this, all the elements belonging to the two equal-regions are relocated to the central region, resulting in the presence of a single equal-region in the center. This region is surrounded by elements of lesser magnitude on the left and elements of greater magnitude on the right.\\
\item {\texttt{The dual-pivot partitioning method}}: The dual-pivot partitioning method was proposed by Vladimir Yaroslavskiy in 2009 \cite{ya} . This method introduced a novel dual-pivot quicksort scheme that demonstrated superior speed compared to alternative methods when dealing with equal elements and randomized data. dual-pivot quicksort used in Oracle’s Java runtime library since version 7. The concept of dual-pivot quicksort involves selecting two pivots of the array, one at the left end and the other at the right end. It is necessary for the left pivot to be as small as or equal to the right pivot. In the left portion, all elements will be smaller than the left pivot. In the middle portion, all elements will be larger than or equal to the left pivot and also will be smaller than or equal to the right pivot and in the right portion, all elements will be larger than or equal to the right pivot. In cases where the left pivot is equal to the right pivot, the elements that are equal to the pivot are positioned in the middle.\\

 \item {\texttt{The Peters partitioning method}}: In 2021, Orson Peters proposed a new method for partitioning and solving the duplicate data problem, which performed better than previous methods \cite{Op}. In this partitioning that is used in pattern-defeating quicksort, It uses two partition functions. Partition-left that elements equal to the pivot are moved to the left part and partition-right that elements equal to the pivot are moved to the right part.\\

\end{itemize}
 All the methods mentioned above to solve the problem of the presence of duplicate data are focused on improving the partitioning method. It's location is marked in the quicksort algorithm.\\

\begin{lstlisting}
   quicksort(int ar[],int start,int end)
    {
 -->  par = partition(ar,start,end)

     quicksort(ar,start,par-1);

     quicksort(ar,par,end);
   }
\end{lstlisting}

 Regrettably, the utilization of these techniques to enhance efficiency and speed up quicksort has resulted in a higher number of algorithm lines and more complexity. In the quest for a streamlined and efficient solution to the aforementioned real data sort problem, a novel partition method were unearthed.\\

\section {Algorithm of eqsort}

  For simpler solution to the problem of duplicate data for quicksort, we explore before partitioning and between two recursive functions.\\

\begin{lstlisting}
   quicksort(int ar[],int start,int end)
    {
 -->
      par = partition(ar,start,end)
      eqsort(ar,start,par-1);
 -->
      eqsort(ar,par,end);
    }
\end{lstlisting}
The first and simplest solution is to use a while loop that skips over the duplicates ( quicksort's missing loop ). In this new approach for handling duplicated element, the method's structure resembles that of the classic quicksort methods(single-pivot 2-way partitioning). We carefully arrange the elements with repetition in their correct positions through multiple steps. The identical elements equal to pivot are grouped together either on the right or left side. Using this approach, the program becomes significantly more concise and does not have the complications of the previous methods like dual-pivot and pdqsort method  (eqsort 22 lines, dual-pivot more than 60 lines). This method is commonly referred to as "equal quicksort" or "eqsort" for short.\\

\textbf{Eqsort1 algorithm in c++ }:

\begin{lstlisting}
1: int partition(int ar[],int start,int end){
2:     int pivot=ar[start];
3:     int i=start,j=end+1;
4:     while(true){
5:          do {i++;} while(ar[i]<pivot);
6:          do {j--;} while(ar[j]>=pivot);
7:          if(i>=j){
8:                   if (ar[j]<pivot) j++;
9:                    if (j>start){
10:                          swap (ar[j-1],ar[start]);
11:                       j--;
                       }
12:                   return j;
            }
13:         swap(ar[i],ar[j]);
        }
     }
14: void eqsort(int ar[],int start,int end){
15:    while (ar[start]==ar[start-1])start++;    /* quicksort's missing loop */
16:    if(start>=end)return;
17:    int par=partition(ar,start,end)
18:    eqsort(ar,start,par-1);
19:    eqsort(ar,par+1,end);
     }
20:  void sort(int ar[],int start,int end){
21:     ar[end+1]=infinite; ar[start-1]=-infinite;
22:     eqsort(ar,start,end);
        }
\end{lstlisting}

\begin{figure*}[h]
  \centering
  \includegraphics[width=15cm , height=8cm]{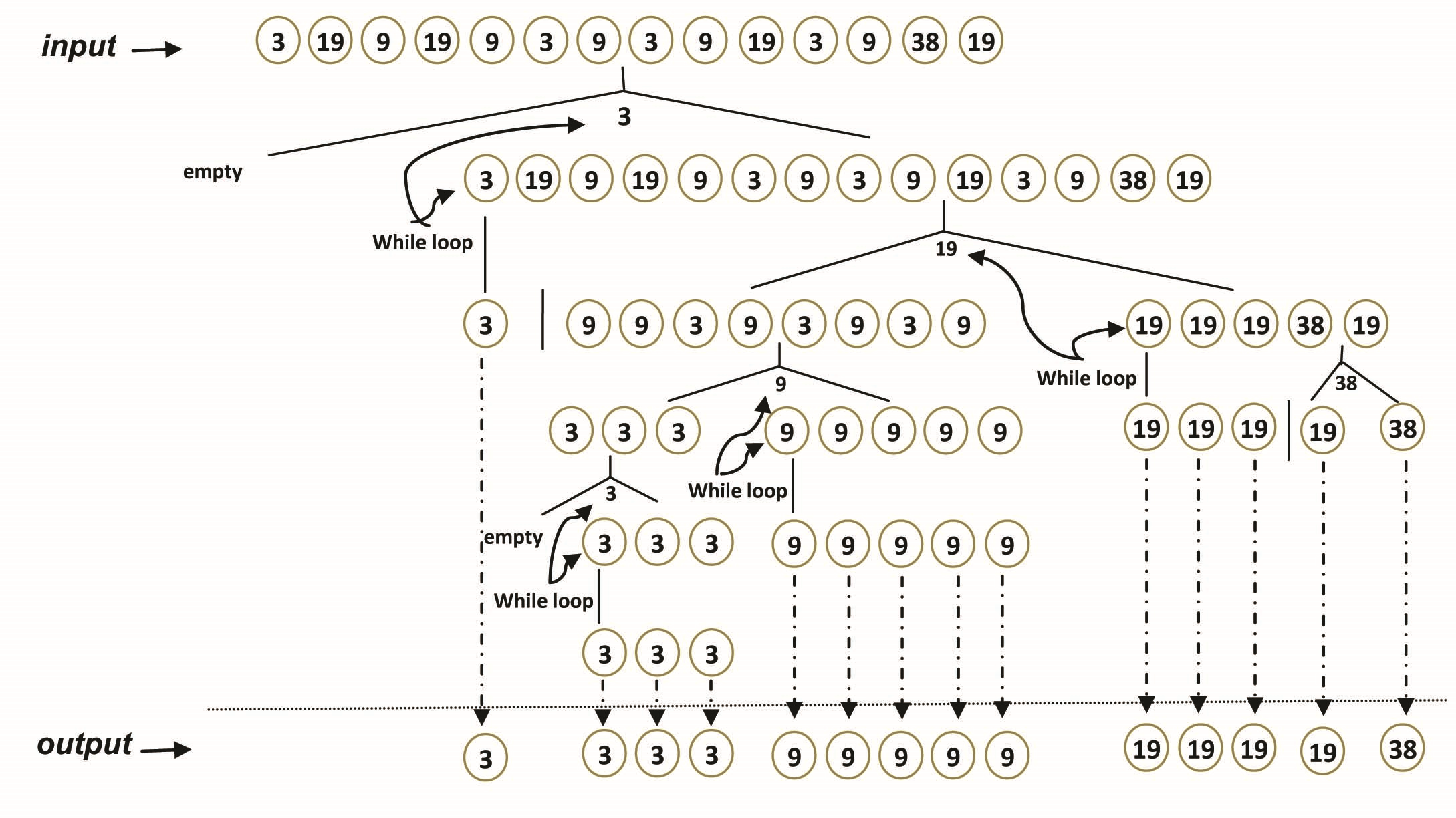}
  \vspace*{-2mm}
  \caption{{{\footnotesize Example of how eqsort2 sorts elements of
  an array. In the example we use the first element as the
  pivot but it could be any other element by swapping it with first element.
  In this example, the first element (3) is selected as the pivot, the right sub-array contains all the elements greater than three and equal to three, and the left sub-array is empty, which is exit in the second step. Before entering the right sub-array, the while loop is executed. Since the first number is 3 and the pivot of this step is 3, it passes through it and for the remaining data, the first element 19 is selected as the pivot and this cycle is repeated.\\}}
}
  \label{fig:illu}
 \end{figure*}

\textbf{Eqsort2 algorithm in c++ }:

\begin{lstlisting}
1: int partition(int ar[],int start,int end){
2:     int pivot=ar[start];
3:     int i=start,j=end+1;
4:     while(true){
5:          do {i++;} while(ar[i]<pivot);
6:          do {j--;} while(ar[j]>=pivot);
7:          if(i>=j){
8:                   if (ar[j]<pivot) j++;
9:                    if (j>start){
10:                          swap (ar[j-1],ar[start]);
11:                       j--;
                       }
12:                   return j;
            }
13:         swap(ar[i],ar[j]);
        }
     }
14: void eqsort(int ar[],int start,int end){
15:    if(start>=end)return;
16:    int par=partition(ar,start,end)
17:    eqsort(ar,start,par-1);
18:    while (ar[par]== ar[par+1]) par++;          /*  quicksort's missing loop */
19:    eqsort(ar,par+1,end);
     }
20:  void sort(int ar[],int start,int end){
21:     ar[end+1]=infinite; ar[start-1]=-infinite;
22:     eqsort(ar,start,end);
        }
\end{lstlisting}
 Figure~\ref{fig:illu} shows how the eqsort2 works. In this method, we can select the first element as a pivot, or select the last element with some minor changes in the algorithm. Obviously, any other selection of pivots to increase its efficiency should be swapped with the first element.

 The second solution is to use an auxiliary partitioning to collect the elements equal to the pivot. We carefully arrange the elements with repetition in their correct positions through one step. \\

 \textbf{Eqsort3 algorithm in c++ }:

\begin{lstlisting}
1: int partition(int ar[],int start,int end){
2:     int pivot=ar[start];
3:     int i=start,j=end+1;
4:     while(true){
5:          do {i++;} while(ar[i]<pivot);
6:          do {j--;} while(ar[j]>=pivot);
7:          if(i>=j){
8:                   if (ar[j]<pivot) j++;
9:                    if (j>start){
10:                          swap (ar[j-1],ar[start]);
11:                       j--;
                       }
12:                   return j;
            }
13:         swap(ar[i],ar[j]);
        }
     }
14:   int partition_aux(int ar,int start,int end){
15:     int pivot=ar[start];
16:     int i=start-1,j=end+1;
17:     while(true){
18:          do {i++;} while(ar[i]==pivot);
19:          do {j--;} while(ar[j]>pivot);
20:          if(i>=j){
21:                   if (ar[j]==pivot) j++;
22:                   return j;
            }
23:         swap(ar[i],ar[j]);
       }
     }
24: void eqsort(int ar[],int start,int end){
25:    if(start<end){
26:       int par = partition(ar,start,end)
27:       eqsort(ar, start, par-1);
28:       par = partition_aux (ar, par,end)
29:       eqsort(ar,par,end);
       }
    }
30: void sort(int ar[],int start,int end){
31:     ar[end+1]=infinite;ar[start-1]=-infinite;
32:     eqsort(ar,start,end);
    }
\end{lstlisting}
We analyze the effectiveness of eqsort1  in comparison to the five well-known methods discussed earlier. The analyzed data of  double  have been chosen, with the potential for extension to similar data.
\section {Experimental setup}
 For the purpose of evaluating the efficiency of the eqsort method in comparison to other methods, we conduct multiple repetitions of sorting various data sets and measure the average  running times, the average comparisons and the average swaps for each method. The dataset used in these evaluations have been chosen at random. Due to the small size of the dataset, the running times is minimal and the results are not very reproducible. For solving this problem, we can utilize the srand function from the c library to generate a multitude of diverse arrays with the possibility of repetition. I would demonstrate the time it takes to sort an array with 1000 data by creating 1000,000 different arrays of 1000 members using each method (n=1000). These one million different arrays have identical properties across all sorting methods. We denoted this number as d( d=1000000). We can determine the average sorting time of these arrays for each method. Reproducibility is extremely high, and conversely, we utilized arrays with varying layouts in this measurement. By increasing d, we try to get a sorting time with 3-digit repeatability so that two methods that have even a 1 percent speed difference can be compared.  We perform the measurement process three times for each method and then calculate the average time  of this three run. For example when the running times are as follows:\\
time1=0.046226 ,   time2=0.046230   ,   time3=0.046242 \\
We utilize the value 0.0462 to facilitate the comparison of these methods.

The partitionings being compared are the six methods mentioned in the previous section. Sort times are displayed for different partitioning methods: T-sedg for Sedgewick partitioning, T-dijk for Dijkstra partitioning, T-bm for Bentley-McIlroy partitioning, T-ydual for Yaroslavskiy partitioning, T-pdqs for the Orson Peters partitioning(pattern-defeating quicksort) and T-eqs for the new partitioning (eqsort).\\
In both the new eqsort method and the five selected methods, none of the speed-up techniques are utilized. This includes not using insertion sort for small data, not choosing a pivot randomly, and not selecting a pivot based on the average of 3 or 9 elements. All of these methods were compared with the original structure.
All algorithms have been implemented in the C++ programming language and have been compiled with the GNU C++ Compiler 11.4.0 .  All tests have been performed on a PC with an i3-380M , 2 core , 2.53 GHz, with 4 GB memory and running Linux Ubuntu 22.04.

\section {Result}
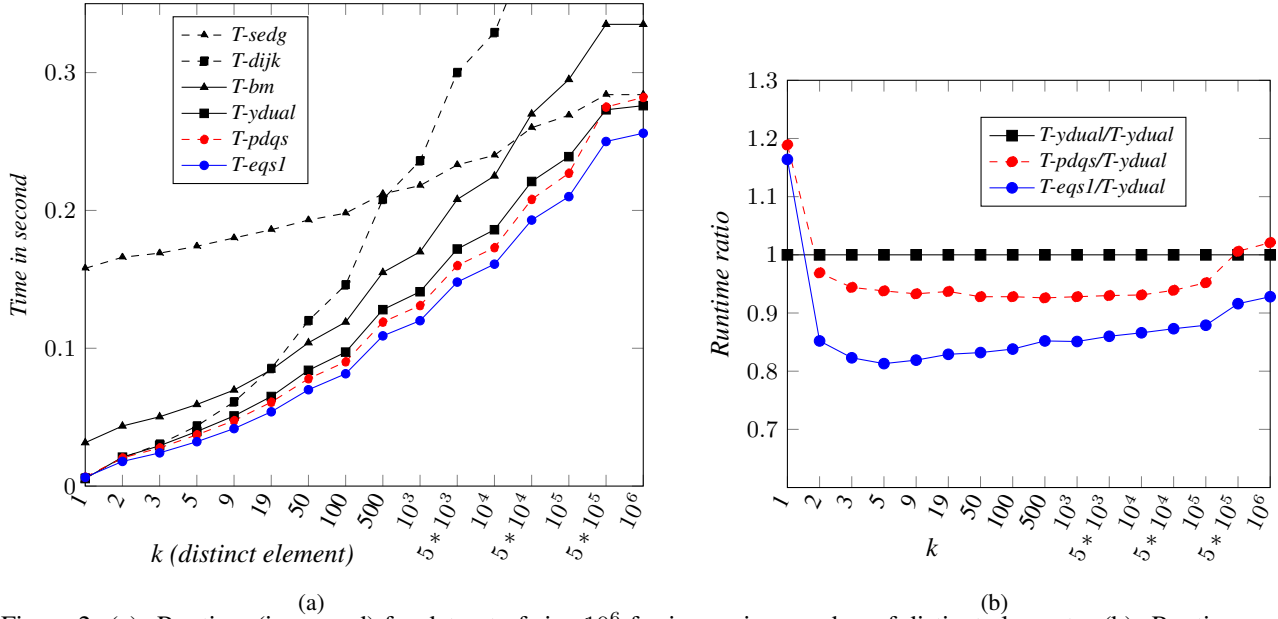
\begin{figure*}[h]
\begin{subfigure}[b]{0.5\linewidth}
\centering
 \begin{tikzpicture}
   \begin{axis}[
   legend style={nodes={scale=0.8},at={(0.40,0.80)},anchor=east},
   legend cell align=left,
   mark options={scale=0.8},
   xmax=16,xmin=1,
   width=9cm,height=8cm,
   ymin= 0,ymax=0.35,
   xticklabel style = {font=\small },
   yticklabel style = {font=\small },
   x label style ={at={(axis description cs:0.3,-0.1)}},
   xlabel=\emph{k (distinct element)}, ylabel=\emph{Time in second},
   ylabel style = {font=\small },
   xtick={1,2,3,4,5,6,7,8,9,10,11,12,13,14,15,16},
   ytick={0,0.1,0.2,...,0.5},
   xticklabels={1,2,3,5,9,19,50,100,500,$10^{3}$,$5*10^{3}$,$10^{4}$,$5*10^{4}$,
   $10^{5}$,$5*10^{5}$,$10^{6}$},
   x tick label style={rotate=63 , anchor=east},
   ]

   \addplot [dashed ,mark=triangle*] coordinates{(1,0.158) (2,0.166) (3,0.169) (4,0.174) (5,0.18 )(6,0.186)(7,0.193 )(8,0.198 )(9,0.212 )(10,0.218 )(11,0.233 )(12,0.24 )(13,0.26 )(14,0.269 )(15,0.284 )(16,0.284 )};
   \addplot[dashed,mark=square*] coordinates{(1,0.00585) (2,0.0205) (3,0.0304) (4,0.0436) (5,0.0611) (6,0.0853)(7,0.120 )(8,0.146 )(9,0.208 )(10,0.236 )(11,0.300)(12,0.329 )(13,0.395 )(14,0.423 )(15,0.487 )(16,0.531 )};
   \addplot[ mark=triangle* ] coordinates{(1,0.0315) (2,0.0436) (3,0.0503) (4,0.0592) (5,0.0697)(6,0.0843)(7,0.104 )(8,0.119 )(9,0.155 )(10,0.17 )(11,0.208 )(12,0.225 )(13,0.27 )(14,0.295 )(15,0.335 )(16,0.335 )};
   \addplot[black ,mark=square*] coordinates{(1,0.00546) (2,0.021) (3,0.0293) (4,0.0396) (5,0.0509)(6,0.065)(7,0.084 )(8,0.0972 )(9,0.128 )(10,0.141 )(11,0.172 )(12,0.186 )(13,0.221 )(14,0.239 )(15,0.273 )(16,0.276 )};
   \addplot [dashed , red,mark=*]coordinates{(1,0.00649) (2,0.0204) (3,0.0277) (4,0.0372) (5,0.0475)(6,0.0609)(7,0.0779 )(8,0.0902 )(9,0.119 )(10,0.131 )(11,0.16 )(12,0.173 )(13,0.208 )(14,0.227 )(15,0.275 )(16,0.282 )};
   \addplot [blue ,mark =*] coordinates{(1, 0.00636) (2,0.0179) (3,0.0241) (4,0.0322) (5,0.0417)(6,0.0539)(7,0.0699 )(8,0.0815 )(9,0.109 )(10,0.120 )(11,0.148 )(12,0.161 )(13,0.193 )(14,0.210 )(15,0.250 )(16,0.256 )};
   \legend{\emph{T-sedg},\emph{T-dijk},\emph{T-bm},\emph{T-ydual},\emph{T-pdqs},\emph{T-eqs1}}
   \end{axis}
   \end{tikzpicture}
   \caption{}
   \end{subfigure}%
   \begin{subfigure}[b]{0.6\linewidth}
   \centering
   \begin{tikzpicture}
   \begin{axis}[
   legend style={nodes={scale=0.8},at={(0.40,0.80)},anchor=west},
   legend cell align=left,
   xmax=16,xmin=1,
   width=8cm,height=7cm,
   ymin= 0.6,ymax=1.3,
   xticklabel style = {font=\small },
   yticklabel style = {font=\small },
   x label style ={at={(axis description cs:0.3,-0.1)}},
   xlabel=\emph{k }, ylabel=\emph{Runtime ratio},
   xtick={1,2,3,4,5,6,7,8,9,10,11,12,13,14,15,16},
   ytick={0.70,0.8,...,1.3},
   xticklabels={1,2,3,5,9,19,50,100,500,$10^{3}$,$5*10^{3}$,$10^{4}$,$5*10^{4}$,
   $10^{5}$,$5*10^{5}$,$10^{6}$},
   x tick label style={rotate=63 , anchor=east},
   ]
   \addplot[black ,mark=square*] coordinates{(1,1) (2,1) (3,1) (4,1) (5,1)(6,1)(7,1) (8,1)(9,1)(10,1)(11,1)(12,1)(13,1)(14,1)(15,1)(16,1 )};
   \addplot [ dashed , red,mark=*] coordinates{(1,1.189) (2,0.969) (3,0.944) (4,0.938) (5,0.933)(6,0.937)(7,0.928)(8,0.928)(9,0.926)(10,0.928)(11,0.930)(12,0.931)
   (13,0.939)(14,0.952)(15,1.006)(16,1.021)};
   \addplot[blue ,mark =*] coordinates{(1,1.164) (2,0.852) (3,0.823) (4,0.813) (5,0.819)(6,0.829)(7,0.832)(8,0.838)(9,0.852)(10,0.851)(11,0.860)(12,0.866)
   (13,0.873)(14,0.879)(15,0.916)(16,0.928)};

   \legend{\emph{T-ydual/T-ydual},\emph{T-pdqs/T-ydual},\emph{T-eqs1/T-ydual}}
   \end{axis}
   \end{tikzpicture}
   \caption{}
   \end{subfigure}
   \vspace*{-7mm}
   \caption{(a): Runtime (in second) for  dataset of size $10^{6}$ for increasing number of distinct elements.
   (b): Runtime ratio for dual-pivot , pdqsort and eqsort methods}
   \label{fig:res6}
   \end{figure*}

Figure~\ref{fig:res6} displays the average sorting times for one million($10^{6}$)  data points across six partitions in double data. For every measurement, 2000 different arrays have been sorted( d=2000) and the times are reported in seconds. number of duplicated are shown with 'k' . First all the elements are chosen uniformly (k=1). Similarly, the number of repetitions is decreased until we ultimately arrive at the array with a minimal amount of repetitions. Most of the numbers selected for k are small to examine the various approaches used in high repetitions.\\

As can be seen in  Fig~\ref{fig:res6}, Sedgewick's method is less dependent on the number of duplicated elements.
Dijkstra-Way Quicksort is fast when there are a lot of duplicates, but starts losing its efficiency when the number of duplicates is smaller. The efficiency of Bentley's method surpasses that of Dijkstra's method when the number of distinct elements exceeds 20. This finding holds true for  datasets with a total of 1000 , 10000 and 100000 elements (Fig~\ref{fig:res5} and Fig~\ref{fig:res4}). In very low repetitions (almost $ k> n/10$), this method exhibits a slower speed in comparison to Sedgewick's method. The dual-pivot , pdqsort and eqsort1 methods exhibit superior performance compared to other methods across all iterations.

 In order to conduct a more thorough analysis and comparison of these three methods, we utilize the data in runtime ratio for dual-pivot, pdqsort and eqsort1. Subsequently, we construct a sort time graph for this methods, as depicted in Figure ~\ref{fig:res6}(b).\\

The diagram depicted in Figure ~\ref{fig:res6}(b) illustrates that under the condition of equal elements, when k = 1, the dual-pivot method exhibits slightly faster. However, as k increases to k = 2, the pdqsort and eqsort1  methods demonstrates accelerated performance. Furthermore, for k=2 to k=0.5n the pdqsort  method demonstrates enhanced speed compare to dual-pivot. The comparison between the pdqsort and eqsort1 shows that the eqsort1 is about 10-15 percent faster than dual-pivot and pdqsort in all amount of k. Result for dataset in size 10000 and 100000 has shown in fig~\ref{fig:res5} and fig~\ref{fig:res4}. Similar results were obtained for integer and string data. The results indicate that the new method  is completely competitive with other methods and works better than them in many cases specially for real-world data .

\section {Comparisons and swaps}
 The average number of swaps and comparisons was measured for three methods in different k. In each measurement, 2000 different arrays of one million data were analyzed. The results show and we also expect (because the complexity of the methods has a direct relationship with k), as the value  of k increases, the amount of swap and comparison increases for all these methods. For a better investigation, we obtain their ratio with the dual–pivot method.
   \begin{figure}[h]
   \begin{center}
   \begin{tikzpicture}[baseline]
   \begin{axis}[
   legend style={nodes={scale=0.6},at={(0.40,0.80)},anchor=east},
   legend cell align=left,
   xmax=15,xmin=1,
   width=9cm,height=7cm,
   ymin= 0.3,ymax=1.5,
   xticklabel style = {font=\small },
   yticklabel style = {font=\small },
   x label style ={at={(axis description cs:0.3,-0.1)}},
   xlabel=\emph{k (distinct element)}, ylabel=\emph{Ratio},
   xtick={1,2,3,4,5,6,7,8,9,10,11,12,13,14,15},
   ytick={0.4,0.5,...,1.4},
   xticklabels={2,3,5,9,19,50,100,500,$10^{3}$,$5*10^{3}$,$10^{4}$,$5*10^{4}$,
   $10^{5}$,$5*10^{5}$,$10^{6}$},
   x tick label style={rotate=63 , anchor=east},
   ]

   \addplot [ black ,mark=square*]coordinates{(1,1) (2,1) (3,1) (4,1) (5,1)(6,1)(7,1) (8,1)(9,1)(10,1)(11,1)(12,1)(13,1)(14,1)(15,1)};
   \addplot [ red,mark=triangle*] coordinates{(1,0.755) (2,0.765) (3,0.781) (4,0.779) (5,0.785)(6,0.775)(7,0.772)(8,0.766)(9,0.760)(10,0.760)(11,0.758)(12,0.757)
   (13,0.755)(14,0.761)(15,0.752)};
   \addplot [blue ,mark =*]coordinates{(1,0.622) (2,0.602) (3,0.596) (4,0.602) (5,0.615)(6,0.631)(7,0.641)(8,0.664)(9,0.668)(10,0.683)(11,0.687)(12,0.702)
   (13,0.710)(14,0.745)(15,0.762)};
    \addplot [dashed ,red , mark=triangle*]coordinates{(1,0.450) (2,0.452) (3,0.458) (4,0.477) (5,0.497
   )(6,0.518)(7,0.510) (8,0.521)(9,0.523)(10,0.526)(11,0.535)(12,0.572)(13,0.558)(14,0.603)(15,0.614)};
   \addplot [ dashed , blue  ,mark=* ]coordinates{(1 , 0.450)(2 , 0.452)(3,0.459) (4,0.477) (5,0.497) (6,0.518)(7,0.510)(8,0.520) (9,0.522)(10,0.530)(11,0.531)(12,0.557)(13,0.531)(14,0.540)(15,0.549)};
   \legend{\emph{com-ydual/com-ydual},\emph{com-pdqs/com-ydual},\emph{com-eqs1/com-ydual},\emph{sw-pdqs/sw-ydual},\emph{sw-eqs1/sw-ydual}}
   \end{axis}
   \end{tikzpicture}
   \vspace*{-2mm}
   \caption{ Comparisons and swaps ratio for dual-pivot , pdqsort and eqsort1 methods.}
   \label{fig:cs}
   \end{center}
   \end{figure}
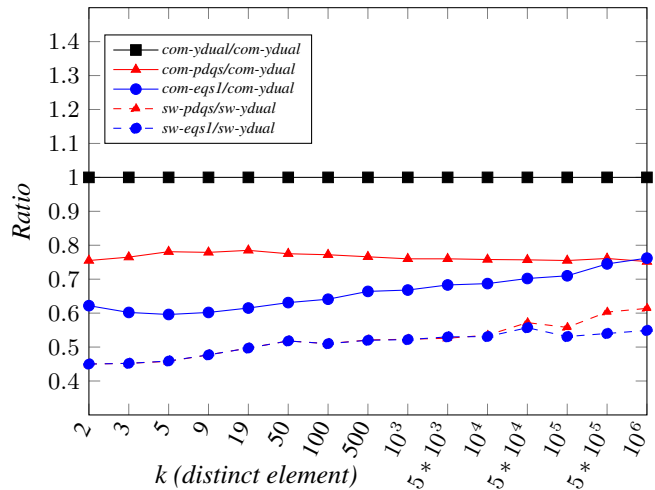

 The results are shown in Fig~\ref{fig:cs}. The eqsort1 method has less number of comparisons  than the pdqsort method  for low amount of K (more duplicates).

  Both method have a lower amount of swaps than the dual-pivot method (about half). The new method performs a lower swap ratio than pdqsort in low duplicated. It seems that the sum of these factors makes eqsort1 method more efficient than other methods.

\section {Conclusion}
This article presents eqsort, a novel and efficient technique for sorting data containing duplicate values. We conducted a comparative experimental analysis of this method with several other widely used methods by evaluating a substantial number of diverse datasets. The experimental findings demonstrated that:\\
\begin{itemize}

\item  Taking into account real-world data, eqsort consistently operates within the high efficiency range. This approach has the potential to supplant the other partitionings.
\end{itemize}
\bibliographystyle{plain}
\bibliography{mybib}
\appendix

\appendix

\begin{figure*}[h]
\begin{subfigure}[b]{0.5\linewidth}
\centering
 \begin{tikzpicture}
   \begin{axis}[
   legend style={nodes={scale=0.8},at={(0.40,0.80)},anchor=east},
   legend cell align=left,
   mark options={scale=0.8},
   xmax=16,xmin=1,
   width=8cm,height=7cm,
   ymin= 0,ymax=0.035,
   xticklabel style = {font=\small },
   yticklabel style = {font=\small },
   x label style ={at={(axis description cs:0.3,-0.1)}},
   xlabel=\emph{k (distinct element)}, ylabel=\emph{Time in second},
   ylabel style = {font=\small },
   xtick={1,2,3,4,5,6,7,8,9,10,11,12,13,14,15,16},
   ytick={0,0.005,0.01,...,0.035},
   xticklabels={1,2,3,5,9,19,50,100,200,500,$10^{3}$,$5*10^{3}$,$10^{4}$,$3*10^{4}$,$5*10^{4}$,$10^{5}$},
   x tick label style={rotate=63 , anchor=east},
   ]

   \addplot [dashed ,mark=triangle*] coordinates{(1,0.0132) (2,0.0143) (3,0.0147) (4,0.0152) (5,0.0157)(6,0.0163)(7,0.0171 )(8,0.0177 )(9,0.0184 )(10,0.0192 )(11,0.0200 )(12,0.0220 )(13,0.0229 )(14,0.0241 )(15,0.0245 )(16,0.0246 )};
   \addplot[dashed,mark=square*] coordinates{(1,0.000556) (2,0.00199) (3,0.00296) (4,0.00426) (5,0.00595)(6,0.00838)(7,0.0118 )(8,0.0144 )(9,0.0170 )(10,0.0206 )(11,0.0233 )(12,0.0298 )(13,0.0327 )(14,0.0372 )(15,0.0391 )(16,0.0409 )};
   \addplot [mark=triangle*]coordinates{(1,0.00307) (2,0.00426) (3,0.00492) (4,0.00579) (5,0.00683)(6,0.00827)(7,0.0103 )(8,0.0118 )(9,0.0133 )(10,0.0154 )(11,0.0171 )(12,0.0217 )(13,0.0241 )(14,0.0274 )(15,0.0281 )(16,0.0284 )};
   \addplot [black ,mark=square*] coordinates{(1,0.000508) (2,0.00204) (3,0.00287) (4,0.00386) (5,0.00500)(6,0.00646)(7,0.00833 )(8,0.00965 )(9,0.0110 )(10,0.0128 )(11,0.0141 )(12,0.0176 )(13,0.0194 )(14,0.0221 )(15,0.0228 )(16,0.0231 )};
   \addplot[dashed , red,mark=*] coordinates{(1,0.000622) (2,0.00198) (3,0.00269) (4,0.00363) (5,0.00467)(6,0.00599)(7,0.00772 )(8,0.00895 )(9,0.0102 )(10,0.0119 )(11,0.0132 )(12,0.0166 )(13,0.0186 )(14,0.0222 )(15,0.0233 )(16 , 0.0241 )};
   \addplot[ blue ,mark =* ] coordinates{(1 , 0.000601) (2 , 0.00174) (3 , 0.00236) (4 , 0.00315) (5 , 0.00408)(6 , 0.00530)(7 , 0.00690 )(8 , 0.00807 )(9 , 0.00926 )(10 , 0.0108 )(11 , 0.0121 )(12 , 0.0153 )(13 , 0.0170)(14 , 0.0201 )(15 , 0.0210 )(16 , 0.0216 )};
   \legend{\emph{T-sedg},\emph{T-dijk},\emph{T-bm},\emph{T-ydual},\emph{T-pdqs},\emph{T-eqs1}}
   \end{axis}
   \end{tikzpicture}
   \caption{}
   \end{subfigure}%
   \begin{subfigure}[b]{0.6\linewidth}
   \centering
   \begin{tikzpicture}
   \begin{axis}[
   legend style={nodes={scale=0.8},at={(0.40,0.80)},anchor=west},
   legend cell align=left,
   xmax=16,xmin=1,
   width=8cm,height=7cm,
   ymin= 0.6,ymax=1.3,
   xticklabel style = {font=\small },
   yticklabel style = {font=\small },
   x label style ={at={(axis description cs:0.3,-0.1)}},
   xlabel=\emph{k }, ylabel=\emph{runtime ratio},
   xtick={1,2,3,4,5,6,7,8,9,10,11,12,13,14,15,16},
   ytick={0.70,0.8,...,1.3},
   xticklabels={1,2,3,5,9,19,50,100,200,500,$10^{3}$,$5*10^{3}$,$10^{4}$,$3*10^{4}$,$5*10^{4}$,$10^{5}$},
   x tick label style={rotate=63 , anchor=east},
   ]
   \addplot[black ,mark=square*] coordinates{(1,1) (2,1) (3,1) (4,1) (5,1)(6,1)(7,1) (8,1)(9,1)(10,1)(11,1)(12,1)(13,1)(14,1)(15,1)(16,1 )};
   \addplot [ dashed , red,mark=*] coordinates{(1,1.2) (2,0.972) (3,0.936) (4,0.940) (5,0.933)(6,0.928)(7,0.927)(8,0.927)(9,0.930)(10,0.929)(11,0.931)(12,0.942)
   (13,0.960)(14,1.004)(15,1.023)(16,1.040)};
   \addplot[dashed ,blue ,mark =star] coordinates{(1,1.017) (2,0.818) (3,0.801) (4,0.804) (5,0.814)(6,0.828)(7,0.851)(8,0.869)(9,0.884)(10,0.901)(11,0.916)(12,0.934)
   (13,0.943)(14,0.981)(15,1.022)(16,1.071)};
   \addplot[ dashed , blue ,mark=triangle*] coordinates{(1,1.16) (2,0.886) (3,0.860) (4,0.857) (5,0.859)(6,0.861)(7,0.865)(8,0.871)(9,0.875)(10,0.876)(11,0.890)(12,0.896)
   (13,0.903)(14,0.927)(15,0.942)(16,0.951)};
   \addplot[ blue ,mark =*] coordinates{(1,1.183) (2,0.852) (3,0.821) (4,0.814) (5,0.816)(6,0.821)(7,0.829)(8,0.836)(9,0.843)(10,0.849)(11,0.858)(12,0.867)
   (13,0.880)(14,0.910)(15,0.921)(16,0.933)};
   \legend{\emph{T-ydual/T-ydual},\emph{T-pdqs/T-ydual},\emph{T-eqs3/T-ydual},\emph{T-eqs2/T-ydual},\emph{T-eqs1/T-ydual}}
   \end{axis}
   \end{tikzpicture}
   \caption{}
   \end{subfigure}
   \vspace*{-7mm}
   \caption{(a): Runtime (in second) for  dataset of size $10^{5}$ for increasing number of distinct elements.
   (b): Runtime ratio for dual-pivot , pdqsort , eqsort1 ,eqsort2 and eqsort3 methods}
   \label{fig:res5}
   \end{figure*}
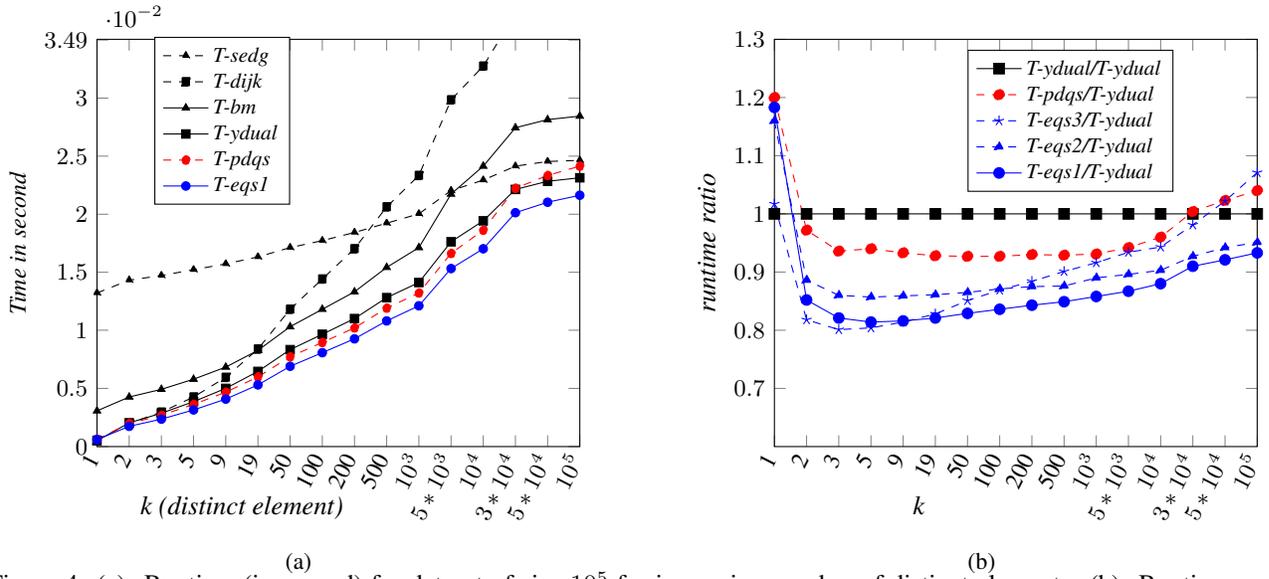

\begin{figure*}[h]
\begin{subfigure}[b]{0.5\linewidth}
\centering
 \begin{tikzpicture}
   \begin{axis}[
   legend style={nodes={scale=0.8},at={(0.40,0.80)},anchor=east},
   legend cell align=left,
   mark options={scale=0.8},
   xmax=16,xmin=1,
   width=8cm,height=7cm,
   ymin= 0,ymax=0.0025,
   xticklabel style = {font=\small },
   yticklabel style = {font=\small },
   x label style ={at={(axis description cs:0.3,-0.1)}},
   xlabel=\emph{k (distinct element)}, ylabel=\emph{Time in second},
   ylabel style = {font=\small },
   xtick={1,2,3,4,5,6,7,8,9,10,11,12,13,14,15,16},
   ytick={0,0.0005,0.001,...,0.0025},
   xticklabels={1,2,3,5,9,19,50,75,100,250,500,750,$10^{3}$,$3*10^{3}$,$5*10^{3}$,$10^{4}$},
   x tick label style={rotate=63 , anchor=east},
   ]
   \addplot [dashed ,mark=triangle*] coordinates{(1,0.00109) (2,0.0012) (3,0.00125) (4,0.00130) (5,0.00135 )(6,0.00142)(7,0.00152 )(8,0.00156 )(9,0.00159 )(10,0.00170 )(11,0.00180 )(12,0.00185 )(13,0.00186 )(14,0.00201 )(15,0.00203 )(16,0.00205 )};
   \addplot[dashed,mark=square*] coordinates{(1,0.000055) (2,0.000198) (3,0.000294) (4,0.000424) (5,0.000594)
   (6,0.000834)(7,0.00118 )(8,0.00133 )(9,0.00144 )(10,0.00180 )(11,0.00208)(12,0.00224 )(13,0.00236 )(14,0.00281 )(15,0.00300 )(16,0.00319 )};
   \addplot[ mark=triangle* ] coordinates{(1,0.000306) (2,0.000425) (3,0.000495) (4,0.000579) (5,0.000685)(6,0.000829)(7,0.00104 )(8,0.00113 )(9,0.00120 )(10,0.00144 )(11,0.00164 )(12,0.00178 )(13,0.00188 )(14,0.00223 )(15,0.00229 )(16,0.00232 )};
   \addplot[black ,mark=square*] coordinates{(1,0.000051) (2,0.000204) (3,0.000285) (4,0.000385) (5,0.000500)(6,0.000646)(7,0.000836 )(8,0.000916 )(9,0.000974 )(10,0.001168 )(11,0.001322 )(12,0.00142 )(13,0.00149 )(14,0.00177 )(15,0.00184 )(16,0.00187 )};
   \addplot [dashed , red,mark=*]coordinates{(1,0.000062) (2,0.000197) (3,0.000270) (4,0.000363) (5,0.000468)(6,0.000604)(7,0.000783 )(8,0.000857 )(9,0.000912 )(10,0.00110 )(11,0.00126 )(12,0.00137 )(13,0.00146 )(14,0.00181 )(15,0.00193 )(16,0.00200 )};
   \addplot [blue ,mark =*] coordinates{(1,0.000067) (2,0.000176) (3,0.000234) (4,0.000315) (5,0.000410)(6,0.000534)(7,0.000700 )(8,0.000780 )(9,0.000823 )(10,0.000996 )(11,0.00114 )(12,0.00124 )(13,0.00133 )(14,0.00163 )(15,0.00171 )(16,0.00177 )};
   \legend{\emph{T-sedg},\emph{T-dijk},\emph{T-bm},\emph{T-ydual},\emph{T-pdqs},\emph{T-eqs1}}
   \end{axis}
   \end{tikzpicture}
   \caption{}
   \end{subfigure}%
   \begin{subfigure}[b]{0.6\linewidth}
   \centering
   \begin{tikzpicture}
   \begin{axis}[
   legend style={nodes={scale=0.8},at={(0.40,0.80)},anchor=west},
   legend cell align=left,
   xmax=16,xmin=1,
   width=8cm,height=7cm,
   ymin= 0.6,ymax=1.3,
   xticklabel style = {font=\small },
   yticklabel style = {font=\small },
   x label style ={at={(axis description cs:0.3,-0.1)}},
   xlabel=\emph{k }, ylabel=\emph{Runtime ratio},
   xtick={1,2,3,4,5,6,7,8,9,10,11,12,13,14,15,16},
   ytick={0.70,0.8,...,1.3},
   xticklabels={1,2,3,5,9,19,50,75,100,250,500,750,$10^{3}$,$3*10^{3}$,$5*10^{3}$,$10^{4}$},
   x tick label style={rotate=63 , anchor=east},
   ]

   \addplot[black ,mark=square*] coordinates{(1,1) (2,1) (3,1) (4,1) (5,1)(6,1)(7,1) (8,1)(9,1)(10,1)(11,1)(12,1)(13,1)(14,1)(15,1)(16,1 )};
   \addplot [ dashed , red,mark=*] coordinates{(1,1.216) (2,0.966) (3,0.947) (4,0.943) (5,0.936)(6,0.935)(7,0.937)(8,0.936)(9,0.936)(10,0.940)(11,0.951)(12,0.962)
   (13,0.974)(14,1.027)(15,1.050)(16,1.067)};
   \addplot[dashed ,blue ,mark =star] coordinates{(1,1) (2,0.814) (3,0.800) (4,0.801) (5,0.814)(6,0.830)(7,0.855)(8,0.868)(9,0.874)(10,0.893)(11,0.911)(12,0.917)
   (13,0.924)(14,0.977)(15,1.025)(16,1.085)};
   \addplot[  blue ,mark=*] coordinates{(1,1.313 ) (2,0.863) (3,0.821) (4,0.818) (5 , 0.820)(6,0.827)(7,0.837)(8,0.842)(9,0.845)(10,0.853)(11,0.865)(12,0.873)(13,0.887)
   (14,0.922)(15,0.933)(16,0.945)};
   \legend{\emph{T-ydual/T-ydual},\emph{T-pdqs/T-ydual},\emph{T-eqs3/T-ydual},\emph{T-eqs1/T-ydual}}
   \end{axis}
   \end{tikzpicture}
   \caption{}
   \end{subfigure}
   \vspace*{-7mm}
   \caption{(a): Runtime (in second) for  dataset of size $10^{4}$ for increasing number of distinct elements.
   (b): Runtime ratio for dual-pivot , pdqsort , eqsort1 and eqsort3 methods}
   \label{fig:res4}
   \end{figure*}
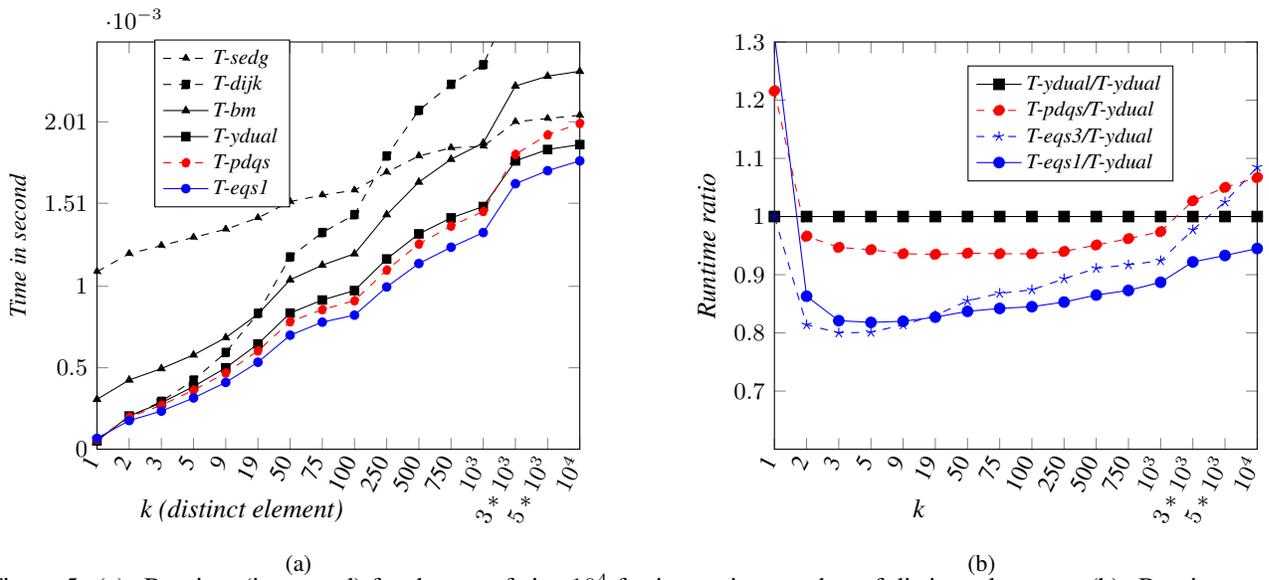

\end{document}